\documentclass[aps,prd,preprint,nofootinbib]{revtex4}
\usepackage{amsfonts}
\usepackage{mathrsfs}
\usepackage{graphicx}
\usepackage{amsmath}
\usepackage{amssymb}
\usepackage{epsfig}
\usepackage{float}
\usepackage{graphicx}
\usepackage{slashed}
\usepackage{color}
\usepackage{multirow}
\usepackage{ulem}
\usepackage{adjustbox}

\usepackage{stmaryrd}
\allowdisplaybreaks[1]

\graphicspath{{fig/}{dia/}} \DeclareGraphicsExtensions{.eps}
\begin{document}

\title{Two-body charmed baryon nonleptonic weak decays with decuplet baryon involving $SU(3)$ breaking effects}

\author {Chao-Qiang Geng$^1$\footnote{cqgeng@ucas.ac.cn}, 
	Chia-Wei Liu$^{2}$\footnote{chiaweiliu@sjtu.edu.cn} and 	
	Sheng-Lin Liu$^{1,3,4}$\footnote{liushenglin22@mails.ucas.ac.cn}}
\affiliation{
	$^1$School of Fundamental Physics and Mathematical Sciences, Hangzhou Institute for Advanced Study,~UCAS, Hangzhou 310024, China\\
	$^2$Tsung-Dao Lee Institute,
	Shanghai Jiao Tong University, Shanghai 200240, China\\
	$^3$Institute of Theoretical Physics,~UCAS, Beijing 100190, China\\
	$^4$University of Chinese Academy of Sciences, 100190 Beijing, China
}
\date{\today}

\begin{abstract}
We study the singly charmed baryon two-body decays of  $\mathbf{B}_c \rightarrow \mathbf{B}_D P$ under $SU(3)_{F}$ flavor symmetry, where ${\bf B}_{c}$ refers to anti-triplet singly charmed baryons, and ${\bf B}_{D}$ and $P$ denote decuplet baryons and pseudoscalar mesons, respectively. 
Compared to the literature, we adopt a   concise and reasonable treatment of the $SU(3)$ flavor symmetry breaking,   by considering the final-state mass dependence of partial waves. With 4 real parameters in total and 9 experimental data points, we achieve a minimal  $\chi^2/\text{d.o.f.} $ value of $ 1.46$, providing a satisfactory explanation to the experiment results. We evaluate the branching ratios and Lee-Yang parameters  suggesting the absence of the strong phase in $\mathbf{B}_c \rightarrow \mathbf{B}_D P$, along with an  unusual positive decay asymmetry $\alpha\,(\Xi_{c}^{0}\rightarrow \Omega^{-}K^{+})$. The predicted results of large $SU(3)_{F}$ breaking effects in $\mathbf{B}_c \rightarrow \mathbf{B}_D P$ can  be verified in the further research by the experiments at BESIII, Belle-II and LHCb.

\end{abstract}

\maketitle

\section{Introduction}
The nonleptonic two-body charmed baryon weak decays of  $\mathbf{B}_c \rightarrow \mathbf{B}_D P$ have been studied extensively by the BESIII~\cite{BESIII:2022udq,BESIII:2018qyg,BESIII:2024mbf,BESIII:2018cvs,BESIII:2023dvx}, Belle-II~\cite{Belle:2020xku}, and LHCb Collaborations~\cite{LHCb:2022sck}, with increasing precision. In this context, $\mathbf{B}_c =  (\Xi^{0}_c, -\Xi^{+}_c, \Lambda^{+}_c)$ represents the anti-triplet charmed baryons, while $\mathbf{B}_D$ and $P$ denote octet baryons and pseudoscalar meson states, respectively. These two-body decays are crucial, as they allow one to extract non-factorizable contributions based on the vanishing of factorizable amplitudes. It provides an excellent opportunity to study non-factorizable effects in singly charmed baryon decays.

Despite the rich experimental data, the calculation of  weak decays for the charm quark into light quarks remains a challenging task. This is primarily due to the large mass of the charm quark $m_c$, which renders the $SU(4)_F$ flavor symmetry ineffective. On the other hand, the heavy quark expansion fails because  $m_c$  is not sufficiently large. The unique decay mechanism of $\mathbf{B}_c \rightarrow \mathbf{B}_D P$ further complicates the application of QCD factorization methods. To overcome these challenges, many approaches for studying charmed hadron decays have been explored~\cite{Cheng:1991sn,Cheng:1995fe,Cheng:1993gf,Zenczykowski:1993hw,Fayyazuddin:1996iy,Dhir:2015tja,Cheng:2018hwl,Hsiao:2020iwc,Cheng:2024rdi,Zhong:2024zme,Zhong:2024qqs,Cheng:2024lsn,Hu:2024uia,Jia:2024pyb}. Among these, the $SU(3)_F$ flavor symmetry method has proven useful in both beauty and charmed hadron decays, with its applicability being demonstrated in both two-body and three-body semileptonic and nonleptonic charmed baryon weak decays~\cite{Savage:1989qr,Savage:1991wu,Pirtskhalava:2011va,Grossman:2012ry,Lu:2016ogy,Geng:2017esc,Geng:2017mxn,Wang:2017gxe,Geng:2018plk,Geng:2018bow,Geng:2018rse,Geng:2018upx,Cen:2019ims,Hsiao:2019yur,Geng:2019awr,Geng:2019bfz,Jia:2019zxi,Geng:2019xbo,Liu:2023dvg,Hsiao:2023mud,Geng:2023pkr,Geng:2024sgq,He:2024pxh}.

In this study, we express the  weak decay amplitudes 
of $\mathbf{B}_c \rightarrow \mathbf{B}_D P$
 in terms of partial waves, specifically $P$-wave and $D$-wave, following a framework similar to the previous work of Ref.~\cite{Geng:2019awr}. Building on a detailed discussion of the non-factorizable contributions under $SU(3)_F$, represented by the irreducible representation $\bar{\mathbf{6}}$, and incorporating new experimental data, we extend the analysis to account for the final-state mass dependency of partial waves and the breaking effects of $SU(3)$ flavor symmetry. This leads to the introduction of 5 real parameters, including one representing the strong phase between the partial waves. Finally, we present the branching ratios and Lee-Yang parameters for  $\mathbf{B}_c \rightarrow \mathbf{B}_D P$, and discuss potential error sources and interesting findings from the new fit.

Our paper is organized as follows: we present the formalism with the explicit amplitudes of all decay channels under the $SU(3)_F$ flavor symmetry and the linear $SU(3)_F$ breaking term
 in Sec.~\ref{Formalism}. In Sec.~\ref{Numerical results}, we present our numerical fitting results and discussions. Our conclusion is given in Sec.~\ref{Conclusions}.

\section{Formalism}\label{Formalism}

The nonleptonic two-body charmed baryon weak decays of $\mathbf{B}_{c}\rightarrow\mathbf{B}_{D}P$ can be proceeded through the Cabibbo-favored~(CF), singly Cabibbo-suppressed~(SCS), and doubly Cabibbo-suppressed~(DCS) charmed quark decays of $c\rightarrow su\bar{d}, c\rightarrow ud\bar{d}\,(us\bar{s})$ and $c\rightarrow du\bar{s}$, respectively. Accordingly, the effective Hamiltonian at tree level is given by~\cite{Buras:1998raa}:
\begin{equation}\label{eq1}
	\mathcal{H}_{eff}=\frac{G_{F}}{\sqrt{2}}\sum_{i=-,+}[V_{c s} V^{*}_{u d}c_{i} O^{ds}_{i}+V_{c d} V^{*}_{u d}c_{i}\,(O^{dd}_{i}-O^{ss}_{i})+V_{c d} V^{*}_{u s}c_{i} O^{sd}_{i}],
\end{equation}
where $G_{F}$ is the  Fermi constant and $c_{i}$ represent the Wilson coefficients, and the axial-vector current four-quark operators are  written as: 
\begin{equation}\label{eq2}
	O^{q_{1}q_{2}}_{\pm}=\frac{1}{2}\left[(\bar{u}q_{1})_{V-A}(\bar{q_{2}}c)_{V-A} \pm(\bar{q_{2}}q_{1})_{V-A}(\bar{u}c)_{V-A}\right].
\end{equation}

The CF, SCS and DCS  modes of charmed quark decays can be written as $c\rightarrow q^{i}q^{j}\,\overline{q}_{k}$ with $q_{i}=(u,d,s)$ is the triplet of light quarks under the $SU(3)_{F}$ flavor symmetry. The form of $q^{i}q^{j}\,\overline{q}_{k}$ can be decomposed as the irreducible representations of $\mathbf{15}$, $\mathbf{\bar{6}}$ and $\mathbf{3}$, in which $\mathbf{15}$ and $\mathbf{\bar{6}}$ correspond to the tree-level color-symmetric operator $O^{q_{1}q_{2}}_{+}$ and the tree-level color-antisymmetric operator $O^{q_{1}q_{2}}_{-}$, respectively. $\mathbf{3}$ represents the negligible loop contribution from the penguin diagrams, suppressed both by $\alpha_{em}^{4}$ and CKM matrix element products. Consequently, the effective Hamiltonian can be divided into the symmetric part $H(\mathbf{15})$ and antisymmetric part of  $H(\bar{\mathbf{6}})$, defined  by 
\begin{equation}
	\mathcal{H}_{eff}= \frac{G_{F}}{\sqrt{2}}\left(c_{+} H(\mathbf{1 5})_{k}^{i j}+c_{-} H(\overline{\mathbf{6}})_{l k}\, \epsilon^{l i j}\right)
	\left(\bar{q}_{i} q^{k}\right)_{V-A}\left(\bar{q}_{j} c\right)_{V-A}.
\end{equation}
Under $SU(3)_{F}$, the three lowest-lying charmed baryon states of $\mathbf{B}_{c}$ form anti-triplet charmed baryon states, and $\mathbf{B}_{D}$ and $P$ belong to decuplet baryon and pseudoscalar meson states. In this work, we adopt the same convention for the $SU(3)_{F}$ tensors as those in previous work~\cite{Geng:2019awr}. Although it is essential to account for the mixing of $\eta$ and $\eta^\prime$, the singlet $\eta^\prime$ will be suppressed under the $SU(3)_F$ flavor symmetry and group theory, and thus its contribution to the physical channels can be neglected.

Unlike the decays  such as $\mathbf{B}_c \rightarrow \mathbf{B}_n P$, the factorizable part  of $\mathbf{B}_c \rightarrow \mathbf{B}_D P$, i.e, $H(\mathbf{15})$, vanishes due to $\left\langle \mathbf{B}_D \right| \bar{q} \gamma_{\mu} (1 - \gamma_5) c \left| \mathbf{B}_c \right\rangle = 0$, as constrained by the $SU(3)_{F}$ symmetry of the light quark pair~\cite{Xu:1992sw,Korner:1970xq,Pati:1970fg,Kohara:1991ug,Chau:1995gk}. Accordingly, the decay amplitude for $\mathbf{B}_c \rightarrow \mathbf{B}_D P$ is given by:
\begin{equation}\label{eq4}
	\mathcal{M}=\left\langle\mathbf{B}_{D} M\right| H(\bar{\mathbf{6}})\left|\mathbf{B}_{c}\right\rangle=i q_{\mu} \bar{w}_{\mathbf{B}_{D}}^{\mu}\left(P-D \gamma_{5}\right) u_{\mathbf{B}_{c}},
\end{equation}
in which $q_{\mu}$ is the four momentum of the pseudoscalar meson $P$, $w_{\mathbf{B}_{D}}^{\mu}$ is the Rarita-Schwinger spinor vector for the spin-$3/2$ particle of the decuplet baryon $\mathbf{B}_{D}$, $P(D)$ refers to the $P(D)$-wave amplitude, and $u_{\mathbf{B}_{c}}$ is the spin-$1/2$ Dirac spinor of $\mathbf{B}_{c}$. Under   $SU(3)_{F}$, $P$ and $D$ can be obtained by the $SU(3)_{F}$ overlapping factor $f_{\mathbf{B}_c\mathbf{B}_D P}$ given by the  $SU(3)$ irreducible-representation amplitude (IRA). As shown in Ref.~\cite{Geng:2019awr}, given 
\begin{equation}
	\begin{aligned}
		&P_{\left(\mathbf{B}_{c} \rightarrow \mathbf{B}_{D} P\right)}=P_{0} f_{\mathbf{B}_{c} \mathbf{B}_{D} P}, \quad D_{\left(\mathbf{B}_{c} \rightarrow \mathbf{B}_{D} P\right)}=D_{0} f_{\mathbf{B}_{c} \mathbf{B}_{D} P},\\
		&f_{\mathbf{B}_{c} \mathbf{B}_{D} P}=\left(\mathbf{B}_{D}^\mathrm{T}\right)_{i j k}\left(\mathbf{B}_{c}\right)_{l} H(\bar{\mathbf{6}})_{o m} (P^\mathrm{T})_{q}^{i} \epsilon^{j l o} \epsilon^{k m q}.
	\end{aligned}
\end{equation}
Consequently, the decay width of $\mathbf{B}_c \rightarrow \mathbf{B}_D P$ can be acquired by:
\begin{equation}\label{eq6}
		\Gamma\left(\mathbf{B}_{c} \rightarrow \mathbf{B}_{D} P\right)=\frac{G_{F}^{2}\,|\vec{q}|}{8 \pi m_{\mathbf{B}_{c}}^{2}}|\langle\overline{\mathcal{M}^{2}}\rangle|=G_{F}^{2}\,\zeta\left(P_{0}^{2}+\xi^{2} D_{0}^{2}\right) f_{\mathbf{B}_{c} \mathbf{B}_{D} P}^{2},		
\end{equation}
while the Lee-Yang parameters have the forms:
\begin{equation}\label{eq7}
\alpha=\frac{2 \xi \operatorname{Re}\left(P_{0} D_{0}^{*}\right)}{|P_{0}|^{2}+\xi^{2}|D_{0}|^{2}},\quad
\beta=\frac{2 \xi \operatorname{Im}\left(P_{0} D_{0}^{*}\right)}{|P_{0}|^{2}+\xi^{2}|D_{0}|^{2}},\quad
\gamma=\frac{-|P_{0}|^{2}+\xi^{2}|D_{0}|^{2}}{|P_{0}|^{2}+\xi^{2}|D_{0}|^{2}},\\
\end{equation}
where $|\vec{q}|$ represents the absolute value of the outgoing three-momentum of   $P$ or  $\mathbf{B}_{D}$, $m_{\mathbf{B}_{c}}$ is the mass of  $\mathbf{B}_{c}$, $|\langle\overline{\mathcal{M}^{2}}\rangle|$ refers to the spin-averaged squared amplitude, $\zeta=\left(E_{\mathbf{B}_{D}}+m_{\mathbf{B}_{D}}\right)|\vec{q}|^{3} m_{\mathbf{B}_{c}} /\left(6 \pi m_{\mathbf{B}_{D}}^{2}\right)$ with $E_{\mathbf{B}_{D}}(m_{\mathbf{B}_{D}})$ stands for the energy (mass) of $\mathbf{B}_{D}$, and $\xi=\sqrt{(E_{\mathbf{B}_{D}}-m_{\mathbf{B}_{D}})/(E_{\mathbf{B}_{D}}+m_{\mathbf{B}_{D}})}$. Without loss of generality, we initially set $P_{0}$ to be positive, and $D_{0}$ to be complex with a phase to $P_{0}$. This phase can be obtained through  a combined measurement of $\alpha$ and $\beta$~\cite{Wang:2024wrm}. In our numerical calculation, we verify this phase is smaller than $10^{-7}$, which has no significant impact on the results~\cite{Hsiao:2020iwc}. Clearly, we drop the phase parameter. As a result, we take the phase parameter to be zero, resulting in  $\beta=0$ for all channels.

Under the exact $SU(3)_{F}$ flavor symmetry, the properties of the $u$, $d$, and $s$ quarks are assumed to be identical, leading to the equal mass (em) scheme, where the masses of the three anti-triplet charmed baryons, the ten decuplet baryons, and the eight pseudoscalar mesons are considered equal. In this scheme, the parameters $\zeta$ and $\xi$ are the same across different decay channels. However, these parameters exhibit considerable differences, as the three-momentum $\vec{q}$ varies significantly in different decays when physical particle masses are used. The typical released energy in ${\bf B}_c \to {\bf B}_D P$ is found to be around $300$~MeV in the same size with the constituent strange quark mass. Hence, the $SU(3)_F$ breaking from $m_s$ is expected to be huge. This leads to the physical mass  (pm) scheme, which incorporates a rough estimate of $SU(3)$ symmetry breaking effects. With the increasing availability of experimental data, both the em and pm schemes exhibit large deviations, with $\chi^2/\text{d.o.f.} = 15.19$ for the em scheme and $\chi^2/\text{d.o.f.} = 10.49$ for the pm one, based on 9 updated data points. The large deviation primarily arises from the channels $\Lambda_{c}^{+} \rightarrow \Xi^{\prime 0} K^{+}$ and $\Xi_{c}^{0} \rightarrow \Omega^{-} K^{+}$, both of which are solely contributed by the $E$-type topological diagrams in the topological diagram amplitude (TDA)~\cite{Hsiao:2020iwc,Wang:2024nxb}, suggesting sizable
 $SU(3)$ breaking effects in these decays. 

Since $P_{0}$ and $D_{0}$ have the same dimensionality as masses, the dependence of these parameters on the masses of the final state particles must be carefully considered. To account for more accurate $SU(3)$ breaking effects beyond the em scheme, we modify $P_{0}$ and $D_{0}$ in Eqs.~(\ref{eq6}) and (\ref{eq7}) into dimensionless parameters $p_{0}$ and $d_{0}$:
\begin{equation}\label{eq8}
	\begin{aligned}
		P_{0}&=p_{0}
		\left[  \bar{m}_{\mathbf{B}_{D}}+\bar{m}_{P}-R_{1}\left(\frac{(M_{\mathbf{B}_{D}}-\bar{m}_{\mathbf{B}_{D}})^{2}}{\bar{m}_{\mathbf{B}_{D}}+\bar{m}_{P}}-R_{1}\frac{(M_{P}-\bar{m}_{P})^{2}}{\bar{m}_{\mathbf{B}_{D}}+\bar{m}_{P}}\right)\right] ,\\
		D_{0}&=d_{0}\left[ \bar{m}_{\mathbf{B}_{D}}+\bar{m}_{P}-R_{2}\left(\frac{(M_{\mathbf{B}_{D}}-\bar{m}_{\mathbf{B}_{D}})^{2}}{\bar{m}_{\mathbf{B}_{D}}+\bar{m}_{P}}-R_{1}\frac{(M_{P}-\bar{m}_{P})^{2}}{\bar{m}_{\mathbf{B}_{D}}+\bar{m}_{P}}\right)\right].
	\end{aligned}
\end{equation}
where $R_{1}$ and $R_{2}$ represent the $SU(3)_{F}$ breaking effects in $P$-wave and $D$-wave amplitudes, $\bar{m}$ stands  for the average masses of hadrons, and the uppercase letter $M$ refers to the  physical masses of hadrons, given by Particle Data Group~(PDG)~\cite{ParticleDataGroup:2024cfk}. We assume that the $SU(3)_{F}$ breaking effects in decuplet baryons and pseudoscalar mesons are equal, as both are generated by $m_{s}-m_{u,d}$. This introduces 4 real parameters in total, to fit with 9 experimental data points from 6 branch ratios and 3 decay asymmetries.
We name this scheme as the modified em (mem) scheme. 

\section{Numerical Results}\label{Numerical results}
\begin{figure}
	\centering
	\includegraphics[width=0.9 \textwidth]{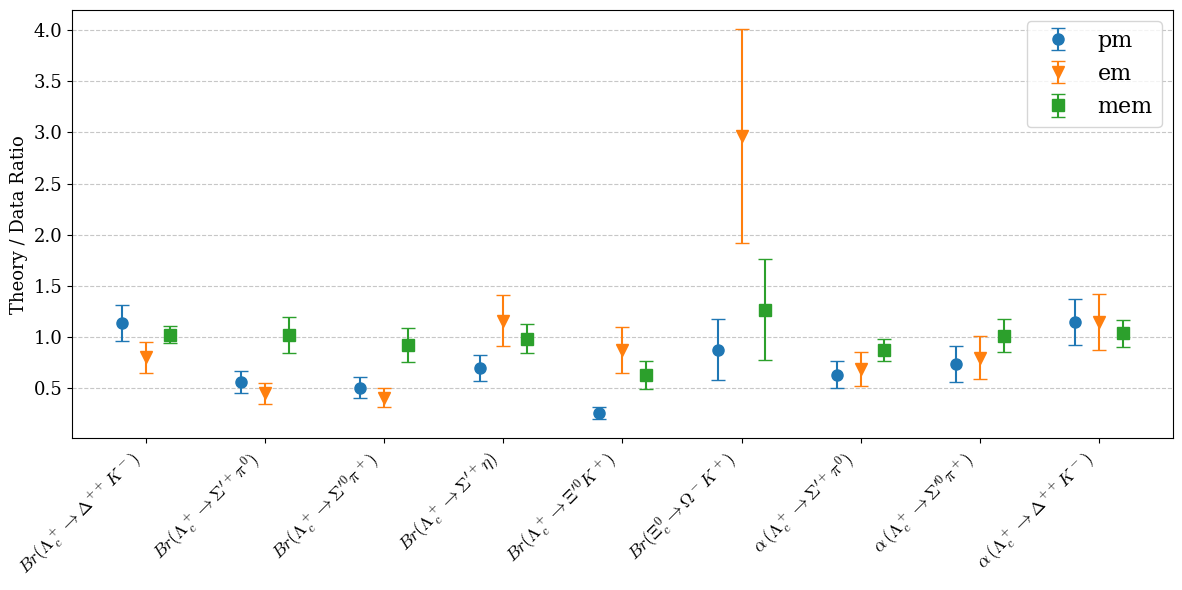}
	\caption{The ratios of theoretical predictions to experimental values for the pm, em, and mem scenarios.}
	\label{fig1}
\end{figure}

In the numerical analysis, we employ the minimum $\chi^{2}$ fitting method to determine the values of the four dimensionless parameters in Eq.~(\ref{eq8}) under $SU(3)_{F}$ for   $\mathbf{B}_{c} \rightarrow \mathbf{B}_{D} P$. The validity of the fit can be tested by the value of  $\chi^{2}/\text{d.o.f.}$. The minimum $\chi^{2}$ fit is given by:
\begin{equation}
	\chi^{2}=\sum_{i}\left(\frac{\mathcal{B}_{S U(3)}^{\,i}-\mathcal{B}_{\text {data }}^{\,i}}{\sigma_{\text {data}}^{\,i}}\right)^{2},
\end{equation}
\begin{table}[H]
	\caption{Predicted branching ratios of  ${\bf B}_{c}\rightarrow {\bf B}_{D}P$.}
	\centering
	\begin{tabular}{ccccc}
		\hline\hline
		Channels&pm~\cite{Geng:2019awr}  &em~\cite{Geng:2019awr} &mem&data.~\cite{ParticleDataGroup:2024cfk,BESIII:2022udq,BESIII:2018qyg,BESIII:2024mbf,BESIII:2018cvs,BESIII:2023dvx,Belle:2020xku,LHCb:2022sck,BaBar:2005uwk}\\
		\hline
		$10^{3}Br\,(\Lambda_{c}^{+}\rightarrow\Delta^{++}K^{-})$&$22.24\pm3.27$&$15.69\pm2.89$&$20.03\pm1.08$&$19.6\pm1.1$\\
		$10^{3}Br\,(\Lambda_{c}^{+}\rightarrow\Sigma^{\prime +}\pi^{0})$&$3.27\pm0.48$&$2.62\pm0.48$&$5.97\pm0.61$&$5.86\pm0.80$\\
		$10^{3}Br\,(\Lambda_{c}^{+}\rightarrow\Sigma^{\prime 0}\pi^{+})$&$3.25\pm0.48$&$2.62\pm0.48$&$5.97\pm0.61$&$6.47\pm0.96$\\
		$10^{3}Br\,(\Lambda_{c}^{+}\rightarrow\Sigma^{\prime +}\eta)$&$4.69\pm0.69$&$7.85\pm1.45$&$6.65\pm0.59$&$6.78\pm0.76$\\
		$10^{3}Br\,(\Lambda_{c}^{+}\rightarrow\Xi^{\prime 0}K^{+})$&$1.54\pm0.23$&$5.23\pm0.96$&$3.75\pm0.42$&$5.99\pm1.09$\\
		$10^{3}Br\,(\Xi_{c}^{0}\rightarrow\Omega^{ -}K^{+})$&$4.64\pm0.70$&$15.69\pm2.89$&$6.71\pm1.67$&$5.3\pm1.6$\\
		$\alpha\,(\Lambda_{c}^{+}\rightarrow\Sigma^{\prime +}\pi^{0})$&$-0.58\pm0.11$&$-0.63\pm0.14$&$-0.80\pm0.06$&$-0.92\pm0.09$\\
		$\alpha\,(\Lambda_{c}^{+}\rightarrow\Sigma^{\prime 0}\pi^{+})$&$-0.58\pm0.11$&$-0.63\pm0.14$&$-0.80\pm0.06$&$-0.79\pm0.11$\\
		$\alpha\,(\Lambda_{c}^{+}\rightarrow\Delta^{+ +}K^{-})$&$-0.63\pm0.11$&$-0.63\pm0.14$&$-0.57\pm0.05$&$-0.55\pm0.05$\\
		\hline\hline
	\end{tabular}
	\label{tab1}
\end{table}
\noindent in which $\mathcal{B}_{S U(3)}^{\,i}$ is the $i-$th decay branching ratio from $SU(3)_{F}$ fitting predictions, $\mathcal{B}_{\text {data}}^{\,i}$ represents the $i-$th experiment data, and $\sigma_{\text {data }}^{\,i}$ stands for the $i-$th experiment error, while $i=1,2,...,9$ for 9 experiment data points in Table.~\ref{tab1}. To compute absolute branching ratios, we incorporate specific branching ratio measurements, $Br\,(\Lambda^{+}_{c}\rightarrow p\pi^{+}K^{-})=(6.84\pm0.34)\%$ and $Br\,(\Xi^{0}_{c}\rightarrow \Xi^{-}\pi^{+})=(1.80\pm0.53)\%$, provided by Belle~\cite{Belle:2013jfq,Belle:2018kzz}. In Table~\ref{tab1}, we present the fitting results for the em, pm, and mem schemes. To examine the discrepancies with the experimental data, we plot the ratio of 
$Br^{\text{scheme}}({\bf B}_c \to {\bf B}_D P)/Br^{\text{data}}({\bf B}_c \to {\bf B}_D P)$ and $\alpha ^{\text{scheme}}({\bf B}_c \to {\bf B}_D P)/\alpha ^{\text{data}}({\bf B}_c \to {\bf B}_D P)$, with schemes including em, pm, and mem, in Fig.~\ref{fig1}. 
The results in the mem scheme give a very good fit with $\chi^2/$d.o.f.$=1.46$, compared to the pm and em schemes.
\begin{table}[H]
	\caption{Fitting values of $p_{0}$, $d_{0}$, $R_{1}$ and $R_{2}$ in unit of 1, and we set $p_{0}$ to be positive.}
	\centering
	\begin{tabular}{cccc}
		\hline\hline
		Parameters&Values$\quad$&Parameters&Values\\
		\hline
		$p_{0}$&$(3.1\pm0.2)\times10^{-2}$$\quad$&$R_{1}$&$3.4\pm0.5$ \\
		$d_{0}$&$-(5.7\pm0.8)\times10^{-2}$$\quad$&$R_{2}$&$8.1\pm1.0$ \\
		\hline\hline
	\end{tabular}
	\label{tab2}
\end{table}
The fitted four parameters are summarized in Table~\ref{tab2}, with the degrees of freedom (d.o.f.) equal to 5. It is noteworthy that these parameters exhibit a $S_{2}$ ambiguity.
This $S_{2}$ ambiguity comes from the exchange of the values of $P_{0}R_{1}$ and $\xi D_{0}R_{2}$, which leads to another set of parameters yielding the same $\chi^{2}/\text{d.o.f.}$ and identical numerical results: $p_{0} = (1.6\pm0.2)\times10^{-2}$, $d_{0} = -(10.5\pm0.6)\times10^{-2}$, $R_{1} = 8.1 \pm 1.0$, and $R_{2} = 3.4 \pm 0.5$. This $S_{2}$ ambiguity can be resolved by considering the contributions from the $P$-wave and $D$-wave parts. 
Since the $D$-wave contributions corresponds to a higher orbital angular momentum $L=2$, it is suppressed in low-momentum regions by the partial wave analysis. As a result, the $P$-wave contribution is expected to be larger than the $D$-wave one. Consequently, we adopt the parameter set in Table~\ref{tab2}, which is consistent with the relation above. In TABLE~\ref{tab3}, we  present the full set of numerical fitting results for the branching ratios of $\mathbf{B}_c \rightarrow \mathbf{B}_D P$.
 
A notable feature of our mem is that since both $R_{1}$ and $R_{2}$ are negative, the channels involving more strange quarks are more strongly suppressed in the branching ratios. Additionally, the isospin symmetry in $ \mathbf{B}_c \rightarrow \mathbf{B}_D P  $ is preserved. In our analysis, the decays of  $\Xi_{c}^{+} \rightarrow \Sigma^{\prime +} \bar{K}^{0}$ and $\Xi_{c}^{+} \rightarrow \Xi^{\prime 0} \pi^{+}$ are still predicted to have zero branching ratios within the IRA  at   tree level under $SU(3)_{F}$, despite the experimental evidence suggesting the existence of these channels~\cite{FOCUS:2003gpe}. These decays are only contributed by the $C^{\prime}$-type topological diagram in the  TDA, which is zero due to the K\"orner-Pati-Woo (KPW) theorem~\cite{Korner:1970xq,Pati:1970fg}. We assume that the observed non-zero and significantly different branching ratios for these channels arise from final-state rescattering (FSR) mechanisms~\cite{Jia:2024pyb} and the linear $SU(3)$ breaking effects in Ref.~\cite{Wang:2024nxb}. However, due to the lack of experimental data points and more precise measurements, these approaches are not included in this work. Nonetheless, precise measurements of the forbidden channels of  $\Xi_{c}^{+} \rightarrow \Sigma^{\prime +} \bar{K}^{0}$ and $\Xi_{c}^{+} \rightarrow \Xi^{\prime 0} \pi^{+}$ are crucial for testing the KPW theorem with the linear $SU(3)$ breaking effects and   FSR mechanism.

The Lee-Yang parameters are summarized in Table~\ref{tab4}, where $\beta = 0$ for all channels. We also justify the upper and lower limits of some of the numerical values to ensure that $ |\alpha| \leq 1 $. Most values of $\alpha\, (B_{c} \rightarrow B_{D} P)$ are negative due to the minus sign between $p_{0}$ and $d_{0}$, except for   $\alpha\, (\Xi_{c}^{0} \rightarrow \Omega^{-} K^{+})$, which exhibits the minimal release energy as well as the maximum $SU(3)$ breaking effects. We encourage experimental collaborations to accumulate more data to verify this decay asymmetry difference in ${\bf B}_{c} \rightarrow {\bf B}_{D} P$. Finally, we update the decay processes of $\Xi_{c}^{0} \rightarrow \Sigma^{\prime 0} K_{S}^{0}/K_{L}^{0}$ in Table.~\ref{tab5}, where $K_{S}^{0} = \frac{1}{\sqrt{2}}(K^{0} + \bar{K}^{0})$ and $K_{L}^{0} = \frac{1}{\sqrt{2}}(K^{0} - \bar{K}^{0})$ are the CP eigenstates. 
\begin{equation}\label{eq11}
	\mathbf{R}  \equiv \frac{\Gamma\left(\Xi_{c}^{0} \rightarrow \Sigma^{\prime 0} K_{S}^{0}\right)-\Gamma\left(\Xi_{c}^{0} \rightarrow \Sigma^{\prime 0} K_{L}^{0}\right)}{\Gamma\left(\Xi_{c}^{0} \rightarrow \Sigma^{\prime 0} K_{S}^{0}\right)+\Gamma\left(\Xi_{c}^{0} \rightarrow \Sigma^{\prime 0} K_{L}^{0}\right)}
	=\frac{\left(c_{c}^{2}+s_{c}^{2}\right)^{2}-\left(c_{c}^{2}-s_{c}^{2}\right)^{2}}{\left(c_{c}^{2}+s_{c}^{2}\right)^{2}+\left(c_{c}^{2}-s_{c}^{2}\right)^{2}}=0.106.
\end{equation}
This $K^{0}_{S}-K^{0}_{L}$ asymmetry in Eq.~(\ref{eq11}) provides a clean prediction in the $SU(3)_{F}$ approach, which can be tested by the experiments in Belle and BESIII.

\begin{table}[H] 
	\caption{Decay branching ratios of ${\bf B}_c\to {\bf B}_D P$ from the reconstruction of the $SU(3)_F$ fit.}
	\renewcommand{\arraystretch}{1}
	\centering
	\begin{tabular}{lclc}
		\hline\hline
		 CF Channels&$\quad$$10^{3}Br$&$\qquad$CF Channels&$\quad$$10^{3}Br$\\
		\hline
		$\Lambda_{c}^{+}\rightarrow\Delta^{++}K^{-}$&$\quad$$20.03\pm1.08$&$\qquad$$\Xi_{c}^{0}\rightarrow\Sigma^{\prime +}K^{-}$&$\quad$$5.79\pm0.35$\\
		$\Lambda_{c}^{+}\rightarrow\Delta^{+}\bar{K}^{0}$&$\quad$$6.56\pm0.36$&$\qquad$$\Xi_{c}^{0}\rightarrow\Sigma^{\prime 0}\bar{K}^{0}$&$\quad$$2.84\pm0.17$\\
		$\Lambda_{c}^{+}\rightarrow\Sigma^{\prime +}\pi^{0}$&$\quad$$5.97\pm0.61$&$\qquad$$\Xi_{c}^{0}\rightarrow\Xi^{\prime 0}\pi^{0}$&$\quad$$4.17\pm0.27$\\
		$\Lambda_{c}^{+}\rightarrow\Sigma^{\prime +}\eta$&$\quad$$6.65\pm0.59$&$\qquad$$\Xi_{c}^{0}\rightarrow\Xi^{\prime 0}\eta$&$\quad$$4.36\pm0.67$\\
		$\Lambda_{c}^{+}\rightarrow\Sigma^{\prime 0}\pi^{+}$&$\quad$$5.97\pm0.61$&$\qquad$$\Xi_{c}^{0}\rightarrow\Xi^{\prime -}\pi^{+}$&$\quad$$8.25\pm0.52$\\
		$\Lambda_{c}^{+}\rightarrow\Xi^{\prime 0}K^{+}$&$\quad$$3.75\pm0.42$&$\qquad$$\Xi_{c}^{0}\rightarrow\Omega^{-}K^{+}$&$\quad$$6.71\pm1.67$\\
		\hline\hline
		SCS Channels&$\quad$$10^{4}Br$&$\qquad$SCS Channels&$\quad$$10^{4}Br$\\
		\hline
		$\Lambda_{c}^{+}\rightarrow\Delta^{++}\pi^{-}$&$\quad$$21.36\pm2.49$&$\qquad$$\Xi_{c}^{0}\rightarrow\Delta^{+}K^{-}$&$\quad$$3.55\pm0.19$\\
		$\Lambda_{c}^{+}\rightarrow\Delta^{+}\pi^{0}$&$\quad$$14.24\pm1.66$&$\qquad$$\Xi_{c}^{0}\rightarrow\Delta^{0}\bar{K}^{0}$&$\quad$$3.49\pm0.19$\\
		$\Lambda_{c}^{+}\rightarrow\Delta^{0}\pi^{+}$&$\quad$$7.12\pm0.83$&$\qquad$$\Xi_{c}^{0}\rightarrow\Sigma^{\prime +}\pi^{-}$&$\quad$$6.36\pm0.65$\\
		$\Lambda_{c}^{+}\rightarrow\Sigma^{\prime +}K^{0}$&$\quad$$3.03\pm0.19$&$\qquad$$\Xi_{c}^{0}\rightarrow\Sigma^{\prime 0}\pi^{0}$&$\quad$$6.35\pm0.65$\\
		$\Lambda_{c}^{+}\rightarrow\Sigma^{\prime 0}K^{+}$&$\quad$$1.54\pm0.09$&$\qquad$$\Xi_{c}^{0}\rightarrow\Sigma^{\prime 0}\eta$&$\quad$$0.78\pm0.07$\\
		$\Xi_{c}^{+}\rightarrow\Delta^{++}K^{-}$&$\quad$$10.67\pm0.58$&$\qquad$$\Xi_{c}^{0}\rightarrow\Sigma^{\prime -}\pi^{+}$&$\quad$$25.26\pm2.58$\\
		$\Xi_{c}^{+}\rightarrow\Delta^{+}\bar{K}^{0}$&$\quad$$3.49\pm0.19$&$\qquad$$\Xi_{c}^{0}\rightarrow\Xi^{\prime 0}K^{0}$&$\quad$$1.96\pm0.23$\\
		$\Xi_{c}^{+}\rightarrow\Sigma^{\prime +}\pi^{0}$&$\quad$$3.18\pm0.33$&$\qquad$$\Xi_{c}^{0}\rightarrow\Xi^{\prime -}K^{+}$&$\quad$$7.89\pm0.90$\\
		$\Xi_{c}^{+}\rightarrow\Sigma^{\prime +}\eta$&$\quad$$3.54\pm0.32$&$\qquad$$\quad$&$\quad$$\quad$\\
		$\Xi_{c}^{+}\rightarrow\Sigma^{\prime 0}\pi^{+}$&$\quad$$3.18\pm0.33$&$\qquad$$\quad$&$\quad$$\quad$\\
		$\Xi_{c}^{+}\rightarrow\Xi^{\prime 0}K^{+}$&$\quad$$2.00\pm0.23$&$\qquad$$\quad$&$\quad$$\quad$\\
		\hline\hline
		DCS Channels&$\quad$$10^{5}Br$&$\qquad$DCS Channels&$\quad$$10^{5}Br$\\
		\hline
		$\Xi_{c}^{+}\rightarrow\Delta^{++}\pi^{-}$&$\quad$$11.37\pm1.32$&$\qquad$$\Xi_{c}^{0}\rightarrow\Delta^{+}\pi^{-}$&$\quad$$3.79\pm0.44$\\
		$\Xi_{c}^{+}\rightarrow\Delta^{+}\pi^{0}$&$\quad$$7.58\pm0.88$&$\qquad$$\Xi_{c}^{0}\rightarrow\Delta^{0}\pi^{0}$&$\quad$$7.58\pm0.88$\\
		$\Xi_{c}^{+}\rightarrow\Delta^{0}\pi^{+}$&$\quad$$3.79\pm0.44$&$\qquad$$\Xi_{c}^{0}\rightarrow\Delta^{-}\pi^{+}$&$\quad$$11.37\pm1.32$\\
		$\Xi_{c}^{+}\rightarrow\Sigma^{\prime +}K^{0}$&$\quad$$1.61\pm0.10$&$\qquad$$\Xi_{c}^{0}\rightarrow\Sigma^{\prime 0}K^{0}$&$\quad$$0.80\pm0.05$\\
		$\Xi_{c}^{+}\rightarrow\Sigma^{\prime 0}K^{+}$&$\quad$$0.82\pm0.05$&$\qquad$$\Xi_{c}^{0}\rightarrow\Sigma^{\prime -}K^{+}$&$\quad$$1.63\pm0.10$\\	
		\hline\hline
	\end{tabular}
	\label{tab3}
\end{table}

\begin{table}[H]
	\caption{Numerical results of the Lee-Yang parameters decays of $\alpha\,(\mathbf{B}_{c}\rightarrow\mathbf{B}_{D}P)$ and $\gamma\,(\mathbf{B}_{c}\rightarrow\mathbf{B}_{D}P)$ in $SU(3)_{F}$.}
	\centering
	\begin{tabular}{lcclcc}
		\hline\hline
		CF Channels&$\,$$\alpha$&$\,\,$$\gamma$&$\qquad$CF Channels&$\,$$\alpha$&$\,\,$$\gamma$\\
		\hline
		$\Lambda_{c}^{+}\rightarrow\Delta^{++}K^{-}$&$\,$$-0.57\pm0.05$&$\,\,$$0.82\pm0.03$&$\qquad$$\Xi_{c}^{0}\rightarrow\Sigma^{\prime +}K^{-}$&$\,$$-0.48\pm0.06$&$\,\,$$0.88\pm0.03$\\
		$\Lambda_{c}^{+}\rightarrow\Delta^{+}\bar{K}^{0}$&$\,$$-0.56\pm0.05$&$\,\,$$0.83\pm0.03$&$\qquad$$\Xi_{c}^{0}\rightarrow\Sigma^{\prime 0}\bar{K}^{0}$&$\,$$-0.46\pm0.07$&$\,\,$$0.89\pm0.03$\\
		$\Lambda_{c}^{+}\rightarrow\Sigma^{\prime +}\pi^{0}$&$\,$$-0.80\pm0.06$&$\,\,$$0.60\pm0.08$&$\qquad$$\Xi_{c}^{0}\rightarrow\Xi^{\prime 0}\pi^{0}$&$\,$$-0.68\pm0.05$&$\,\,$$0.74\pm0.04$\\
		$\Lambda_{c}^{+}\rightarrow\Sigma^{\prime +}\eta$&$\,$$-0.26\pm0.13$&$\,\,$$0.97^{+0.03}_{-0.04}$&$\qquad$$\Xi_{c}^{0}\rightarrow\Xi^{\prime 0}\eta$&$\,$$0.27\pm0.31$&$\,\,$$0.96^{+0.04}_{-0.08}$\\
		$\Lambda_{c}^{+}\rightarrow\Sigma^{\prime 0}\pi^{+}$&$\,$$-0.80\pm0.06$&$\,\,$$0.60\pm0.08$&$\qquad$$\Xi_{c}^{0}\rightarrow\Xi^{\prime -}\pi^{+}$&$\,$$-0.67\pm0.04$&$\,\,$$0.74\pm0.04$\\
		$\Lambda_{c}^{+}\rightarrow\Xi^{\prime 0}K^{+}$&$\,$$-0.09\pm0.20$&$\,\,$$1.00^{+0.00}_{-0.02}$&$\qquad$$\quad$&$\,$$\quad$&$\,\,$$\quad$\\
		\hline\hline
		SCS Channels&$\,$$\alpha$&$\,\,$$\gamma$&$\qquad$SCS Channels&$\,$$\alpha$&$\,\,$$\gamma$\\
		\hline
		$\Lambda_{c}^{+}\rightarrow\Delta^{++}\pi^{-}$&$\,$$-0.83\pm0.06$&$\,\,$$0.55\pm0.09$&$\qquad$$\Xi_{c}^{0}\rightarrow\Delta^{+}K^{-}$&$\,$$-0.57\pm0.05$&$\,\,$$0.82\pm0.03$\\
		$\Lambda_{c}^{+}\rightarrow\Delta^{+}\pi^{0}$&$\,$$-0.83\pm0.06$&$\,\,$$0.55\pm0.09$&$\qquad$$\Xi_{c}^{0}\rightarrow\Delta^{0}\bar{K}^{0}$&$\,$$-0.56\pm0.05$&$\,\,$$0.83\pm0.03$\\
		$\Lambda_{c}^{+}\rightarrow\Delta^{0}\pi^{+}$&$\,$$-0.83\pm0.06$&$\,\,$$0.55\pm0.09$&$\qquad$$\Xi_{c}^{0}\rightarrow\Sigma^{\prime +}\pi^{-}$&$\,$$-0.80\pm0.06$&$\,\,$$0.60\pm0.08$\\
		$\Lambda_{c}^{+}\rightarrow\Sigma^{\prime +}K^{0}$&$\,$$-0.46\pm0.07$&$\,\,$$0.89\pm0.03$&$\qquad$$\Xi_{c}^{0}\rightarrow\Sigma^{\prime 0}\pi^{0}$&$\,$$-0.80\pm0.06$&$\,\,$$0.60\pm0.08$\\
		$\Lambda_{c}^{+}\rightarrow\Sigma^{\prime 0}K^{+}$&$\,$$-0.48\pm0.06$&$\,\,$$0.88\pm0.03$&$\qquad$$\Xi_{c}^{0}\rightarrow\Sigma^{\prime 0}\eta$&$\,$$-0.26\pm0.13$&$\,\,$$0.97^{+0.03}_{-0.04}$\\
		$\Xi_{c}^{+}\rightarrow\Delta^{++}K^{-}$&$\,$$-0.57\pm0.05$&$\,\,$$0.82\pm0.03$&$\qquad$$\Xi_{c}^{0}\rightarrow\Sigma^{\prime -}\pi^{+}$&$\,$$-0.80\pm0.06$&$\,\,$$0.60\pm0.08$\\
		$\Xi_{c}^{+}\rightarrow\Delta^{+}\bar{K}^{0}$&$\,$$-0.56\pm0.05$&$\,\,$$0.83\pm0.03$&$\qquad$$\Xi_{c}^{0}\rightarrow\Xi^{\prime 0}K^{0}$&$\,$$-0.07\pm0.20$&$\,\,$$1.00^{+0.00}_{-0.01}$\\
		$\Xi_{c}^{+}\rightarrow\Sigma^{\prime +}\pi^{0}$&$\,$$-0.80\pm0.06$&$\,\,$$0.60\pm0.08$&$\qquad$$\Xi_{c}^{0}\rightarrow\Xi^{\prime -}K^{+}$&$\,$$-0.07\pm0.20$&$\,\,$$1.00^{+0.00}_{-0.01}$\\
		$\Xi_{c}^{+}\rightarrow\Sigma^{\prime +}\eta$&$\,$$-0.26\pm0.13$&$\,\,$$0.97^{+0.03}_{-0.04}$&$\qquad$$\quad$&$\quad$&$\quad$$\quad$\\
		$\Xi_{c}^{+}\rightarrow\Sigma^{\prime 0}\pi^{+}$&$\,$$-0.80\pm0.06$&$\,\,$$0.60\pm0.08$&$\qquad$$\quad$&$\quad$&$\quad$$\quad$\\
		$\Xi_{c}^{+}\rightarrow\Xi^{\prime 0}K^{+}$&$\,$$-0.09\pm0.20$&$\,\,$$1.00^{+0.00}_{-0.02}$&$\qquad$$\quad$&$\quad$&$\quad$$\quad$\\
		\hline\hline
		DCS Channels&$\,$$\alpha$&$\,\,$$\gamma$&$\qquad$DCS Channels&$\,$$\alpha$&$\,\,$$\gamma$\\
		\hline
		$\Xi_{c}^{+}\rightarrow\Delta^{++}\pi^{-}$&$\,$$-0.83\pm0.06$&$\,\,$$0.55\pm0.09$&$\qquad$$\Xi_{c}^{0}\rightarrow\Delta^{+}\pi^{-}$&$\,$$-0.83\pm0.06$&$\,\,$$0.55\pm0.09$\\
		$\Xi_{c}^{+}\rightarrow\Delta^{+}\pi^{0}$&$\,$$-0.83\pm0.06$&$\,\,$$0.55\pm0.09$&$\qquad$$\Xi_{c}^{0}\rightarrow\Delta^{0}\pi^{0}$&$\,$$-0.83\pm0.06$&$\,\,$$0.55\pm0.09$\\
		$\Xi_{c}^{+}\rightarrow\Delta^{0}\pi^{+}$&$\,$$-0.83\pm0.06$&$\,\,$$0.55\pm0.09$&$\qquad$$\Xi_{c}^{0}\rightarrow\Delta^{-}\pi^{+}$&$\,$$-0.83\pm0.06$&$\,\,$$0.55\pm0.09$\\
		$\Xi_{c}^{+}\rightarrow\Sigma^{\prime +}K^{0}$&$\,$$-0.46\pm0.07$&$\,\,$$0.89\pm0.03$&$\qquad$$\Xi_{c}^{0}\rightarrow\Sigma^{\prime 0}K^{0}$&$\,$$-0.46\pm0.07$&$\,\,$$0.89\pm0.03$\\
		$\Xi_{c}^{+}\rightarrow\Sigma^{\prime 0}K^{+}$&$\,$$-0.48\pm0.06$&$\,\,$$0.88\pm0.03$&$\qquad$$\Xi_{c}^{0}\rightarrow\Sigma^{\prime -}K^{+}$&$\,$$-0.47\pm0.06$&$\,\,$$0.88\pm0.03$\\	
		\hline\hline
	\end{tabular}
	\label{tab4}
\end{table}

\begin{table}[H]
	\caption{Results for $\Xi_{c}^{0}\rightarrow\Sigma^{\prime 0}K_{S}^{0}/K_{L}^{0}$ in $SU(3)_{F}$ with $SU(3)$ breaking effects}
	\centering
	\begin{tabular}{cccc}
		\hline\hline
		channels&$\quad$$f_{\mathbf{B}_{c} \mathbf{B}_{D} P}$&$\quad$$\alpha$&$\quad$$10^{3}Br$\\
		\hline
		$\Xi_{c}^{0}\rightarrow\Sigma^{\prime 0}K_{S}^{0}$&$\quad$$\frac{\sqrt{3}}{3}\sin^{2}\theta+\frac{\sqrt{3}}{3}\cos^{2}\theta$&$\quad$$-0.46\pm0.07$&$\quad$$1.57\pm0.01$ \\
		$\Xi_{c}^{0}\rightarrow\Sigma^{\prime 0}K_{L}^{0}$&$\quad$$\frac{\sqrt{3}}{3}\sin^{2}\theta-\frac{\sqrt{3}}{3}\cos^{2}\theta$&$\quad$$-0.46\pm0.07$&$\quad$$1.27\pm0.01$ \\
		\hline\hline
	\end{tabular}
	\label{tab5}
\end{table}

\section{Conclusions}\label{Conclusions}
We have studied the decay branching ratios and Lee-Yang parameters in the charmed baryon weak decays of $\mathbf{B}_{c}\rightarrow\mathbf{B}_{D}P$ under the $SU(3)_{F}$ flavor symmetry. In these decays, there involve only   non-factorizable effects of $H(\mathbf{\bar{6}})$, and   one-diagram contributions both in IRA and TDA. In contrast to the previous work, where the  em  and  pm  schemes were employed, we have considered the mem scheme with  the dependence on the final-state masses of the $P(D)$-wave, incorporating $ SU(3) $ symmetry breaking effects. Assuming equal $ SU(3) $ breaking effects for decuplet baryons and pseudoscalar mesons, the fitting results are in good agreement with the experimental data, yielding   $ \chi^2/\text{d.o.f} = 1.46 $. The $ S_{2} $ ambiguity in the parameter space of the minimal $ \chi^2 $ fit is resolved by recognizing that the $ P $-wave contribution is dominant over the $ D $-wave one. In particular, we have demonstrated that the strong phase between the $ P $-wave and $ D $-wave contributions is negligible, and the isospin symmetry is preserved despite the inclusion of $ SU(3) $ breaking effects. For the Lee-Yang parameters, most of the values of $ \alpha\, ( \mathbf{B}_{c} \rightarrow \mathbf{B}_{D} P ) $ are negative due to the minus sign between $ p_{0} $ and $ d_{0} $, with the exception of $ \alpha\, (\Xi_{c}^{0} \rightarrow \Omega^{-} K^{+}) $, which is positive due to  the significant $ SU(3) $ breaking in this channel. This particular result should be tested experimentally to probe the   $ SU(3) $ symmetry breaking. Moreover, the forbidden decays of  $ \Xi_{c}^{+} \rightarrow \Sigma^{\prime +} \bar{K}^{0} $ and $ \Xi_{c}^{+} \rightarrow \Xi^{\prime 0} \pi^{+} $ hold substantial significance in verifying the validity of the KPW  theorem, as well as the FSR  mechanism and linear $ SU(3) $ breaking effects in $ \mathbf{B}_{c} \rightarrow \mathbf{B}_{D} P $. Additionally, we have presented the updated branching ratios for the decays of  $ \Xi_{c}^{0} \rightarrow \Sigma^{\prime 0} K_{S}^{0}/K_{L}^{0} $, and calculated the fitting-independent $ K_{S}-K_{L} $ asymmetry, $ \mathbf{R}\, (\Xi_{c}^{0} \rightarrow \Sigma^{\prime 0} K_{S}^{0}/K_{L}^{0}) = 0.106 $. This asymmetry represents a clear prediction of the $ SU(3)_{F} $ framework for charmed baryon decays, which should be of great interest for experimental measurements at Belle, BESIII, and LHCb.

\section*{Acknowledgments} 
 This work is supported in part by the National Key Research and Development Program of China under Grant No. 2020YFC2201501 and  the National Natural Science Foundation of China (NSFC) under Grant No. 12347103 and 12205063.

\end{document}